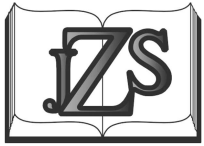

# Classification of Brainwave Signals Based on Hybrid Deep Learning and an Evolutionary Algorithm

Zhyar Rzgar K. Rostam[1], Sozan Abdullah Mahmood[1]

*1 Department of Computer, College of Science, University of Sulaimani*
*E-mail: zhyar.rostam@univsul.edu.iq*
*E-mail: sozan.mahmood@univsul.edu.iq*

**Introduction**

Human brain produces a lot of signals, these signals are non-stationary, and they can be measured via different devices, i.e. functional magnetic resonance imaging (fMRI), magnetoencephalography (MEG), and electroencephalogram (EEG). The brain signal has five waves alpha, beta, delta, gamma and theta. Nowadays, EEG is used in many scientific fields such as neurology (for diagnosing many neurological disorders), education, computer games, and natural language processing. EEG is an easy and an inexpensive way compared to other methods [1].

There are two different kinds of EEG sensor: dry electrodes and gel-based. Previously, EEG sensors were costly; hence their use was limited to laboratories and hospitals [2].

To achieve successful classification, studies mainly focus on pre-processing and feature extraction techniques, because the nature of brain signals is complex and have nonlinearity. Therefore, advanced analysis methods are required for classifying brainwave signals. Deep learning and EAs are among successful techniques that are used in image processing and computer games [3], Natural Language Processing (NLP) [4], and mental state detection [5]. Genetic algorithms (GA) are optimization techniques, or can be classified as a heuristic search technique to find the optimal solution(s). GAs belongs to the Evolutionary Algorithms (EA) that use techniques inspired by evolutionary biology such as: inheritance, mutation, selection, and crossover [6].

Zulkifley et al (2019) [7], proposed a model for classifying brainwave signal, the signals are classified into two groups which are: new task, or routine task. The CNN has been used to achieve the correct classification. The presented CNN architecture is made up of four layers of CNN and three layers of fully connected layers, and RecLU is used as an activation function in each layer. Adam optimizer with a cross entropy loss function is used to train the network. As a result the model achieved accuracy 79.56%.

Rajendra et al (2018) [8], presented a model for detection and diagnosis of seizure using EEG signals based on deep CNN. The dataset used in the study is collected by Andrzejak et al [9]. Before the data is fed to a (1D-CNN) one-dimensional convolutional neural network the signal is normalized, then the normalized data is fed to a 13-layer deep CNN to detect normal, preictal, and seizure classes. The first 10 layers consist of a pair of convolution and Max-pooling layer with different size, consecutively. The final 3 layers are dense layers. The accuracy achieved with proposed model is 88.67%.

Yıldırım et al (2018) [1], proposed a model to identify abnormal EEG signals based on deep convolutional neural network, researchers in [1] used the TUH EEG abnormal corpus (v2.0.0) and they have proposed a new 1D-CNN model to build automated identification of abnormal EEG signals. Eventually, the model is able to detect the abnormal EEG signals with rate of 79.34% and 79.64% accuracy and precision, respectively.

Tang et al (2017) [10], have shown a method that depends on deep CNN for single trail motor imagery EEG. To classify MI tasks (left hand and right hand movement) a model is designed that consists of five-layer CNN, and then the experimental dataset are fed to the CNN. The dataset is prepared by recording data from 28 active electrodes. A comparison has been done between the result obtained from this model and other conventional classification methods such as (power + Support Vector Machine (SVM), Common Spatial Pattern (CSP) + SVM, and Auto-Regressive (AR) + SVM). They conclude that the CNN can improve classification performance; the average accuracy reached using CNN is $86.41 \pm 0.77$.

Furthermore, Uktveris and et al (2017) [11], have proposed a model based on CNN for classifying four-class motor imagery problem. Eleven different CNN architectures are investigated, starting from the simplest architecture and ending with more complex one. A CNN with learning rate 0.01, momentum 0.1, batch size 128, epoch 200, input tensor (22x22x1), and convolution filters (4x4, 16) was trained and tested for final evaluation on all subjects. This architecture achieves best accuracy, which is 70% for training and 68% for testing.

The above mentioned models are all derived from CNN architecture, all of them follow the same basic layers of CNN, i.e. convolution, pooling, and fully connected layers. However, many of the models have used several pooling layers to get the most interesting features. Some of the models have used multiple fully connected layers to achieve better accuracy. Nevertheless, the accuracy results are below 90%.

The main goal of the present study is obtain the best classification results with the highest accuracy rate by combining a deep learning approach (CNN) with evolutionary algorithm (EA) to classify brainwave signals. Instead of using several pooling layers, a single Discrete Wavelet Transform (DWT) Coiflet filter have been used for feature extraction. The model is tested for color and shape identification. In order to achieve this, a dataset is developed by recording EEG signals sourced from six participants and it is used to test the proposed CNN model.

**Materials and Methods**
 *A. Age-Layered Population Structure (ALPS) Genetic Algorithm (GA)*

ALPS is a method that is used to reduce a common problem which occurs during running an EA known as premature convergence when numbers of evolutions the population has done converge to local optima, and no improvements occur regardless of how many times the EA is run. The difference between ALPS and any other EA is that it uses multiple layers for each population. ALPS measures how long the genetic materials have been evolving in the population: offspring age starts from 1 plus age of their oldest parent instead of starting from 0 as in the traditional measurement of age. This helps to randomly generate individuals in the youngest layer [12] [13].

An ALPS-EA works as follows. The algorithm starts by configuring the age layers and then creating, and evaluating, an initial, random population. Once the initial population is created and evaluated, ALPS-EA enters its main loop which consists of cycling through the layers, from $L_{n-1}$ to $L_0$, (Fig.1) and then evolving the EA in that layer for one generation.

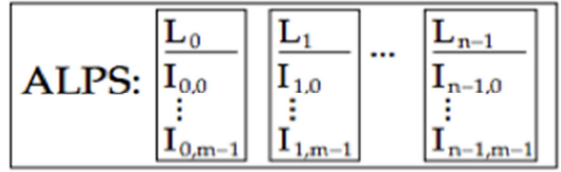

**Fig. 1**. The layout of an ALPS system with n layers ($L_0$ to $L_{n-1}$) and m individuals in each layer ($I_{i,0}$ to $I_{i,m-1}$) [13].

*B. Convolutional Neural Network*

One of the common deep artificial networks is convolutional neural network (CNN). This model takes a major role, especially within image classification. It can identify and classify objects reliably. There are three main layers available with standard CNN architecture which are: convolution layer, pooling layer, and fully connected layers (dense layers). The CNN architecture receives input data with a volume (tensor) form which has width, height, and depth. Filtering and feature extraction have been done in both convolutional and pooling layers, and a fully connected layer (dense layer) is used for classification [14]. The convolution process occurs between the input data and filter(s) (kernel(s)), then an active function is applied on each convolved volume commonly rectified linear unit (RecLU) which is used as an activation function. The pooling layer acts as a feature extractor applied on the results received from the convolution layer. Finally, flatten is a process applied on the extracted feature then these features are fed to the fully connected layer to perform a classification process [15] (Fig. 2).

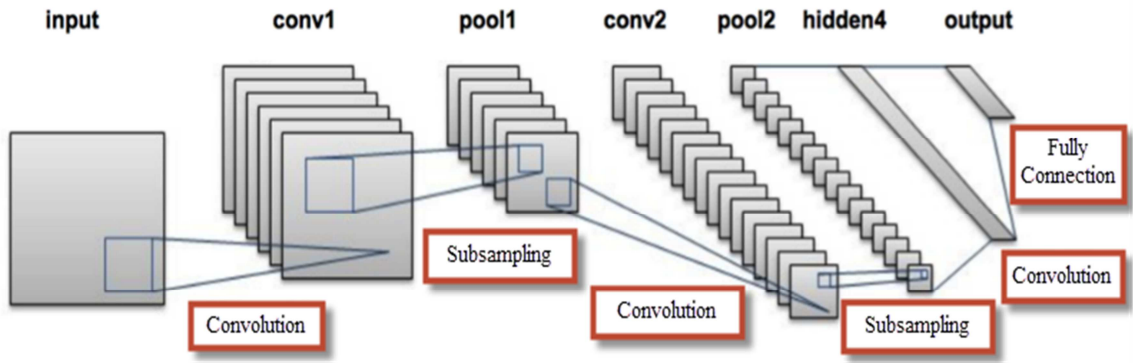

Fig. 2. An overview of a convolutional neural network (CNN) architecture [16].

**Data Collection and Preparation**

In this study the latest version of NeuroSky mind wave which is mindwave mobile 2 has been chosen. The device transfers different EEG signals, i.e. raw signals or raw data which are the main source of information received at 512 Hz, power spectrum (alpha, beta, delta, theta and gamma), attention and meditation level, and eye blink detection [17]. Collecting brainwave signals from six health subjects has created a dataset, each has normal color vision, with a normal mental state and age ranging from 30 ± 5. To provide a good interface for establishing communication with the EEG device, and recording EEG signals according to the requirements, a complete program has been developed. The application can successfully communicate with the EEG device, and then receive the signals transmitted from the EEG device. The brain activities can also be monitored through the application. The raw EEG data, all brainwave bands (alpha, beta, delta, gamma, and theta), attention, meditation, signal quality, and blinking strength are plotted. All the data have been received once a second except the raw EEG data that was received with rate 512 Hz [18].

Each participant was seated on a comfortable chair in a dark room faced with a 43 inch screen. The distance between the subjects and the screen was 125 centimeters. The screen was set to a normal mode, which has a normal brightness with normal color mode, and contrast.

There are two main categories of brainwaves available in the dataset which are colors and shapes. Each category contains two sub-categories (visible and invisible mode). Five sessions of recording data were available in each sub-category, and each session had two sub-sessions, the duration of each sub-session was 25 seconds. Red and green colors used for both visible and invisible mode dataset creation, and forward and right arrow were used for shape dataset creation. These two figures below (Fig. 3, and 4) show two samples of the raw EEG of 100 records, for invisible color green, and visible shape forward.

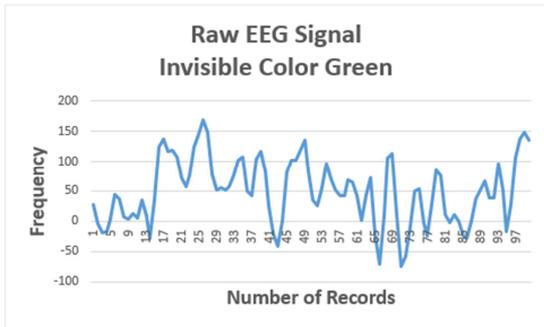
Fig. 3. Raw EEG (Invisible - Green)

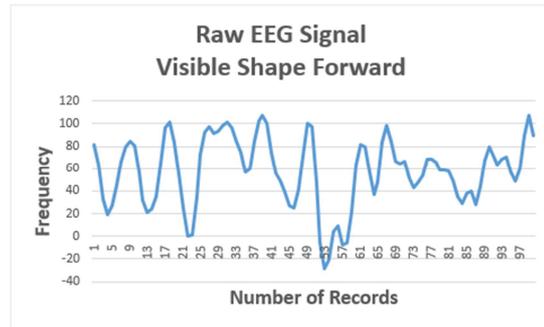
Fig. 4. Raw EEG (Visible - Forward)

The total duration was calculated for all sub-sessions in each mode (visible and invisible) individually. Number of sessions are 5 and duration is 25 seconds. Since this is, also total time duration in each mode are calculated. As a result, the dataset contains 6000 seconds of brainwave signals, which are equal to 100 minutes. Each sub-session duration is 750 seconds. In each mode there are two (colors or shapes) available, total duration for each mode for both colors and shapes is 1500 seconds individually. Thus, the duration of each mode is 3000 seconds.

The common format used for storing data in the dataset is Comma-separated values (CSV). In this study, a separate (CSV) file has been created for each sub-session within individual subjects. Twenty five (25) seconds of raw data have been recorded in each sub-session. Thus, each file contains approximately 506 records per second [18]. To record the real and accurate brainwave signals, the subject's concentration is compulsory. The data preparation in this study involves three main steps: first, from each sub-session, only 10 seconds have been fetched, based on the maximum attention with respect to the order of recording data. Second, within each second only 500 records have been fetched. In rare cases there have been less than 5000 records per 10 seconds. To overcome this problem, a neighbor interpolation has been done until the required size is reached. Finally, the fetched data was merged into a single CSV file according to their categories with respect to their mode.

The number of subjects is 6, the number of records per 10 seconds is 5000, and the number of session is 5, these values are required to calculate the total sub-session records in each mode. The total sub-session records in each mode is 150,000 records. The algorithm for preparing this data is shown below:

***Begin***
*Declare path of csv file to read, attention_list, Maximum_attention_list_time, result_list*
***attention_list*** ← *read 10 maximum attention value from **csv** file*

    ***For*** *i in attention_list **do**:*

    ***Maximum_attention_list_time*** ← *Fetch and map the **maximum 10 attention** values with their time*

    ***End For***

    ***For*** *data in Maximum_attention_list_time **do**:*

   *read RAWEEG from .CSV file*

   *if Length of RAWEEG >= 500*

     *result_list ← fetch top 500 records*

   *else if Length of RAWEEG < 500*

     *result_list ← do neighbor Interpolation for RAWEEG until = 500*

  *End For*

*Write result_list in(.csv) file*

*End.*

### Proposed Model Architecture

  For classifying brainwave signals with high accuracy rate, this study proposed a new CNN model, the proposed model contains three main steps which are: convolution, discrete wavelet transform (DWT) coiflet, and classification. In this model two main changes have been applied to the standard CNN.

  The first modification has been done in the second layer (pooling layer), DWT coiflet 1 has been applied in this layer instead of max or average pooling. DWT Coiflet is derived from daubechies wavelet and member of the discrete wavelet transform families. The wavelet families are used for feature extraction [19], see equation 1, 2 for discrete wavelet transform (DWT) function [20].

$$W_\emptyset(j_0, k) = \frac{1}{\sqrt{M}} \sum_x f(x) \emptyset_{j_0,k}(x) \quad (1)$$

$$W_\Psi(j, k) = \frac{1}{\sqrt{M}} \sum_x f(x) \Psi_{j,k}(x) \quad (2)$$

  The second modification was replacing the fully connected (dense) layer with ALPS-GA [12] and symbolic discriminant analysis (SDA). SDA can determine optimal functional form and coefficients of the discriminate function whether linear or nonlinear [21]. SDA originally was inspired by symbolic regression. The aim of SDA is to overcome limitations of linear discriminate analysis (LDA) which is used for signal classification [22].

  In the convolution layer, convolution process has been applied between the input data (raw EEG) and kernels (filters), then RecLU active function has been applied on each convolved feature. The output generated from this layer has been used as input to the DWT coiflet layer. In the DWT coiflet layer the most interesting features are extracted and the size of received features is reduced by 2. Finally, extracted features have been fed to the ALPS with the SDA to predict the output.

  A standard CNN accepts input data in the form of volume (tensor), in this study the input data is raw EEG, and the raw EEG consists of single raw data. Thus, the input data for the proposed model is set to 1, 5000, 1 for width, height, and depth respectively. The number of kernels used with this model is set to 3 kernels, each kernel size is set to 1, 10 based on the best result achieved. The architecture of the proposed model presented in Fig. 3, all modifications in the model are marked with orange color.

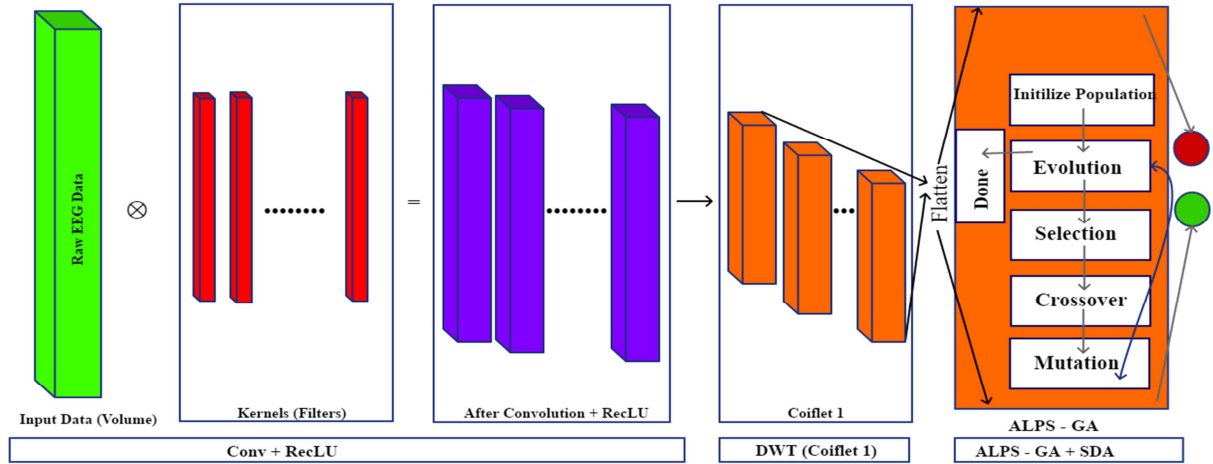

Fig. 5. Proposed model architecture

The following pseudo code explains the algorithm of proposed model which is applied on the dataset.

***Start***
*Declare variables RawEEGData, Kernels, Convolved, Rectified, Coiflet, MaxGeneration and Counter*
*Initialize **Kernels** randomly between (0 and 1); **Counter** ← 0*
  *For all RawEEGData list ∈ dataset do*
      *Convolved ← Calculate convolution between (kernels and list)*
  *End for*
 *For all Convolved features ∈ Convolved do*
     *Rectified ← Math.Max(0, features)*
 *End for*
 *For all Rectified row ∈ Rectified do*
     *Coiflet ← Apply coiflet of order 1 on each row.*
*End for*
*While (Counter < MaxGeneration) do*
   *For all Coiflet row ∈ Coiflet do*
      *Fed each row of the Coiflet use ALPS-GA, and SDA as classifier*
      *Compute Mean Square Error*
   *End for*
   *Counter ← Counter + 1*
*End While*
*End*

**Result and Discussion**
    All results and experiments with a comparison between standard CNN and the proposed model have been provided in this section. The experiments show that the proposed model achieves the highest accuracy. The best result achieved in the visible color mode for accuracy is 92%, precision is 100%, recall is 83%, and F-Measure is 90%.
    Both standard CNN and proposed CNN have been applied on the dataset, the input volume for each model is set to 10 seconds. The dataset has been divided to a training part and a testing part. In order to

achieve the best result, different ratios (percentages) have been used, i.e. %70 - %30, %75-%25, and 80%-%20 for training and testing, respectively. The best result in both models reached in case of %80-%20.

Kernels have been tested with different sizes and different number of kernels. The structure of kernel size is written in this format: width, height and count (number of kernels), both models have been trained using the kernel architectures such as (1x10, 3), (1x8, 3), (1x6x3), (1x8x4)… etc.

The standard CNN involves three main steps, which are: convolution, pooling, and fully connected layer. The first step is the same as the proposed CNN. The convolution process between kernels (filters) and input data has been carried out, but in the standard CNN, max pooling has been applied instead of coiflet. The output of this step is flattened and directly fed to the fully connected layer. Finally, softmax has been used to predict the output classes.

The highest score has been reached with the standard CNN in visible arrow. The results are: accuracy 83%, precision 75%, recall 100%, and F-Measure 85%.

In order to evaluate the proposed model the accuracy, precision, recall and F-measure are calculated according to the equations below:

$$Accuracy = \frac{TN+TP}{TP+FP+TN+FN} \quad (3)$$

$$Recall = \frac{TP}{TP+FN} \quad (4)$$

$$Precision = \frac{TP}{TP+FP} \quad (5)$$

$$F-measure = \frac{2*recall*precision}{recall+precision} \quad (6)$$

The result of comparison between the standard CNN and the proposed model is presented in table 1.

Table 1. Comparison between the standard CNN and the proposed model
Batch size = 1, 5000, 1;   Kernel size = 1, 10;   Kernel count = 3

| Dataset categories | CNN | Generation / L.R | Population Size / Epoch | Accuracy | Mutation probability | MSE/ Loss | Precision | Recall | F-Measure |
|---|---|---|---|---|---|---|---|---|---|
| Visible Color | **Proposed** | 702 | 300 | 0.92 | 0.18 | 0.07 | 1 | 0.83 | 0.9 |
| | **Standard** | 0.005 | 2000 | 0.5 | - | 0.01 | 0.5 | 0.5 | 0.5 |
| Invisible Color | **Proposed** | 384 | 250 | 0.83 | 0.18 | 0.1 | 0.83 | 0.83 | 0.83 |
| | **Standard** | 0.012 | 4500 | 0.58 | - | 0.00002 | 0.57 | 0.66 | 0.61 |
| Visible Arrow | **Proposed** | 804 | 300 | 0.83 | 0.18 | 0.1 | 0.83 | 0.83 | 0.83 |
| | **Standard** | 0.004 | 2000 | 0.83 | - | 0.005 | 0.75 | 1 | 0.857 |
| Invisible Arrow | **Proposed** | 640 | 200 | 0.75 | 0.15 | 0.2 | 0.8 | 0.66 | 0.72 |
| | **Standard** | 0.0035 | 6500 | 0.5 | - | 0.001 | 0.5 | 0.66 | 0.568 |

The dataset for the testing section contains 30,000 records for each sub-session individually. The confusion matrices of the proposed model for visible and invisible colors are presented in table 2. Table 3 provides the confusion matrices for visible and invisible shapes.

Table 2.  Confusion matrices for visible and invisible color (proposed CNN)

| | Actual Red | Actual Green |
|---|---|---|
| Predict Red (visible) | 25000 (TP) | 0 (FP) |
| Predict Green (visible) | 5000 (FN) | 30000 (TN) |
| Predict Red (invisible) | 25000 (TP) | 5000 (FP) |
| Predict Green (invisible) | 5000 (FN) | 25000 (TN) |

Table 3. Confusion matrices for visible and invisible shape (proposed CNN)

|  | *Actual Forward* | *Actual Right* |
|---|---|---|
| *Predict Forward (visible)* | 25000 (TP) | 5000 (FP) |
| *Predict Right (visible)* | 5000 (FN) | 25000 (TN) |
| *Predict Forward (invisible)* | 20000 (TP) | 5000 (FP) |
| *Predict Right (invisible)* | 10000 (FN) | 25000 (TN) |

The tables below show the confusion matrices for both colors and shapes when the standard CNN has been applied on the dataset. The visible and invisible colors are presented in table 4. Table 5 shows the confusion matrices for visible and invisible shapes.

Table 4. Confusion matrices for visible and invisible color (standard CNN)

|  | *Actual Red* | *Actual Green* |
|---|---|---|
| *Predict Red (visible)* | 15000 (TP) | 15000 (FP) |
| *Predict Green (visible)* | 15000 (FN) | 15000 (TN) |
| *Predict Red (invisible)* | 20000 (TP) | 15000 (FP) |
| *Predict Green (invisible)* | 10000 (FN) | 15000 (TN) |

Table 5. Confusion matrices for visible and invisible shape (standard CNN)

|  | *Actual Forward* | *Actual Right* |
|---|---|---|
| *Predict Forward (visible)* | 30000 (TP) | 10000 (FP) |
| *Predict Right (visible)* | 0 (FN) | 20000 (TN) |
| *Predict Forward (invisible)* | 20000 (TP) | 20000 (FP) |
| *Predict Right (invisible)* | 10000 (FN) | 10000 (TN) |

Finally, in table 6 a comparison has been provided between all previous works and the proposed CNN architecture, it shows that the present CNN achieves best accuracy compared to other models.

Table 6. Comparison between previous works and proposed CNN Architecture

| *No* | *References* | *Method* | *Conv. Layer* | *Pool. Layer* | *FC. Layer* | *Application* | *Best Accuracy* |
|---|---|---|---|---|---|---|---|
| 1 | *Zulkifley et al (2019)* [7] | *Based on CNN* | 4 | 1 | 3 | *Classify brain task (new, or routine)* | 79.56% |
| 2 | *Rajendra et al (2018)* [8] | *Based on CNN* | 5 | 5 | 3 | *Detection seizure* | 88.67% |
| 3 | *Yıldırım et al (2018)* [1] | *Based on CNN* | 10 | 5 | 1 | *Identification abnormal EEG signal* | 79.34% |
| 4 | *Tang et al (2017)* [10] | *Modified CNN* | 2 | 0 | 1 | *Classify two motor imagery task* | 86.41 ± 0.77 |
| 5 | *Uktveris et al (2017)* [11] | *Modified CNN* | 1 | 1 | 1 | *Classify four motor imagery task* | 68% |
| 6 | **Proposed CNN Architecture** | *Hybrid CNN and EA* | 1 | 0, (DWT Coiflet | 0, ALPS-GA | *Color, arrow classification* | 92% |

**Conclusion**

A robust brain computer interface (BCI) application requires a model that can classify brainwave signals in an efficient way and interpret these waves to an action. The BCI applications can used in various domains such as: entertainment, gaming, and helping paralyzed people to perform their daily tasks. It is a challenging task to classify brainwave signals from a single electrode. Nowadays, however, the best result with a highest accuracy can be achieved by using deep learning methods. In this study, a new CNN model has been proposed, then a comparison between the standard CNN and the proposed model has been done in order to identify which one has the ability to achieve better results. The results show that the proposed model gives a higher accuracy when both models are applied on the same dataset. The proposed model can reach the best classification when applied on a visible color mode, in this case the result for accuracy is 92%, precision is 100%, and recall is 83%. However, for invisible color and visible shape, the result is 83% for accuracy, precision, and recall. Finally, the minimum result that the proposed model produces is when it is applied on invisible shapes, the result for accuracy is 75%, precision is 80%, and recall is 66%.

**Acknowledgment**

The authors would like to thank the editorial office of the journal for reviewing the manuscript. Furthermore, the authors would like to thank all the people assisted in data collection.